# Carbon Nanotube Based Magnetic Tunnel Junctions

H. Mehrez[1], Jeremy Taylor[1], Hong Guo[1], Jian Wang[2], and Christopher Roland[3]
*1. Center for the Physics of Materials and Department of Physics, McGill University, Montreal, PQ, Canada H3A 2T8.*
*2. Department of Physics, The University of Hong Kong, Pokfulam Road, Hong Kong, China.*
*3. Department of Physics, The North Carolina State University, Raleigh, NC USA 27695.*

Spin-coherent quantum transport in carbon nanotube magnetic tunnel junctions was investigated theoretically. A spin-valve effect is found for metallic, armchair tubes, with a magneto-conductance ratio ranging up to 20%. Because of the finite length of the nanotube junctions, transport is dominated by resonant transmission. The magnetic tunnel junctions are found to have distinctly different transport behavior depending on whether or not the length of the tubes is commensurate with a $3N+1$ rule, with $N$ the number of basic carbon repeat units along the nanotube length.

In a recent experiment, Tsukagoshi, Alphenaar and Ago[1] fabricated molecular scale spin-polarized tunnel junctions by ferromagnetically contacting a carbon nanotube. The data showed that nanotubes have very long spin-scattering lengths of at least 130nm. Similarly, earlier studies showed that carbon nanotubes behave as ballistic quantum conductors with long phase-coherence lengths for the charge carriers[2,3]. All of these facts indicate that nanotubes may well be ideal candidates for achieving molecular scale magnetoelectronics[4], in which *both the charge and spin* degrees of freedom are utilized for the operation of a functional device. While Ref. 1 represents the first measurement to date of such a nanotube-based magneto-transport device, one can expect many more to follow in the near future. Although there have been many contributions concerning quantum transport through carbon nanotubes[5–9], there is currently no theory for their magnetoelectronics. Moreover, conventional magnetic tunnel junction theory[10] cannot be expected to work, as it does not directly incorporate the molecular electronic properties of the constituent material. In order to fill this gap, we have theoretically explored spin-polarized transport through single-wall carbon nanotubes (SWCN), both with and without defects.

The spin-valve effect reported in Ref. 1 is due to the misalignment of the magnetic moments of the two electrodes connecting the nanotubes. This misalignment, which gives rise to a hysteretic magnetoresistence, is expected even if the same ferromagnetic (FM) material is used for the two electrodes contacting the nanotubes. Because of the small size of the nanotube diameter, only a small number of the magnetic domains of the electrodes will be in contact with the nanotube, each of which is likely to have a different orientation of its local magnetic moment[1], even though the average moment of each of the electrodes will be the same. In contrast, conventional tunneling magnetoresistance (TMR)[11] of a magnetic multilayer is nonzero when the multilayer uses materials of different coercivities. This fact alone makes SWCN interesting from a device physics point of view.

To explore spin-polarized transport through carbon nanotubes, we investigated the tunnel junction shown in Fig. (1a). It consists of a SWCN connected to two electrodes whose FM moments $\mathbf{M}$ points in different directions, thereby forming a $FM/SWCN/FM$ device. For simplicity we assume $\mathbf{M}$ of the left (L) electrode points to the $z$-direction, while that of the right (R) electrode points at an angle $\theta$ away from $z$-direction in the $x-z$ plane. Our main results show that this device has a clear spin-valve effect, so that the resistance varies smoothly with angle $\theta$, giving rise to a magneto-conductance ratio up to about 20%. In distinct contrast to the more familiar quantized conductance steps of infinitely long SWCN[5–9] the system is dominated by resonance transmission which is sensitively dependent on whether or not the nanotube length is commensurate with a $3N+1$ rule, where $N$ is the number of basic carbon repeat units along the nanotube[12].

To analyze the SWCN magnetic tunnel junction, we combined the nonequilibrium Green's function technique[13,14] with a simple tight-binding model for the nanotubes. Standard but tedious algebra shows[13,14] that the zero temperature and zero bias spin dependent conductance is given by ($\hbar = 1$):

$$G = \frac{2e^2}{\pi}\sum_{\sigma} Tr\left[Im(\mathbf{\Sigma}^r_L)\mathbf{G}^r Im(\mathbf{\Sigma}^r_R)\mathbf{G}^a\right]_{\sigma\sigma} \ , \tag{1}$$

where subscript $\alpha = L, R$ indicates either the left or right electrode; the spin index $\sigma = -\bar{\sigma}$ takes values $\pm 1$ (or $\uparrow, \downarrow$); and the trace is over the state index[14]. Here, $\mathbf{G}^{r,a}$ indicates the $2n \times 2n$ matrix ($n$ is the number of atoms in the nanotube and 2 is due to spin indices) for the retarded or advanced Green's function, respectively. The self-energy, which describes the coupling of the nanotube to the electrodes, is found to be given by:

$$\mathbf{\Sigma}^r_\alpha(E) = \hat{R}_\alpha \begin{pmatrix} \Sigma^r_{\alpha\uparrow} & 0 \\ 0 & \Sigma^r_{\alpha\downarrow} \end{pmatrix} \hat{R}^\dagger_\alpha, \tag{2}$$

with the rotational matrix $\hat{R}_\alpha$ for lead $\alpha$ defined as

$$\hat{R} = \begin{pmatrix} \cos\theta_\alpha/2 & \sin\theta_\alpha/2 \\ -\sin\theta_\alpha/2 & \cos\theta_\alpha/2 \end{pmatrix}, \tag{3}$$

where $\theta_L \equiv 0$ and $\theta_R \equiv \theta$ for the device shown in Fig. (1a). Expressions (1-3) form the basis of all our subsequent numerical calculations.

In this formalism, all the properties of the magnetic electrodes are contained in the self-energy $\Sigma^r_{\alpha\sigma}$ of (2), which we have parameterized by the corresponding line width function $\Gamma_{\alpha\sigma} = -2Im(\Sigma^r_{\alpha\sigma})$. The retarded (advanced) Green's function of the device is calculated directly using a tight-binding model via

$$\mathbf{G}^r(E) = \frac{1}{E - H_{tube} - \mathbf{\Sigma}^r} \tag{4}$$

with $\mathbf{\Sigma}^r \equiv \mathbf{\Sigma}^r_L(E - qV_L) + \mathbf{\Sigma}^r_R(E - qV_R)$. The nanotube Hamiltonian $H_{tube}$ is modeled using a nearest-neighbor $\pi-$orbital tight binding model with bond potential $V_{pp\pi} = -2.75$ eV. This model is known to give a reasonable, qualitative description of the electronic and transport properties of carbon nanotubes[6,8]. Finally, using the well known Julliere model[10], we fix[15] $\Gamma_{\alpha\uparrow}/\Gamma_{\alpha\downarrow} = 2.0$, corresponding to the commonly used wideband limit of Green's function theory[13].

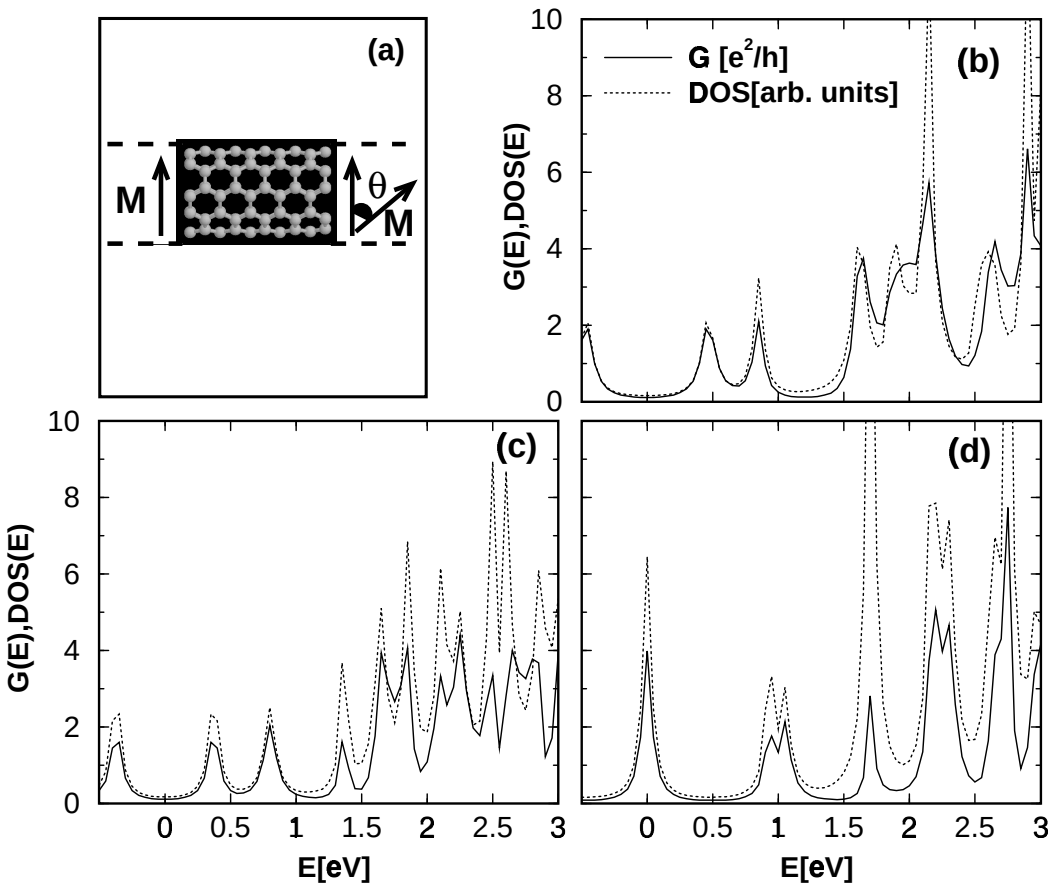

FIG. 1. Conductance (solid lines - units $e^2/h$) and LDOS (dashed lines) for (5,5) SWCN magnetic tunnel device with different lengths as a funtion of the energy $E$ (eV), with $E_F$ fixed at 0.0 eV, $\Gamma_\uparrow = 0.6$ and $\Gamma_\downarrow = 0.3$. (b) for length $N = 5$; (c) for $N = 6$; and (d) for $N = 7$. A schematic plot of the single wall carbon nanotube magnetic tunnel juntion is given in (a), with FM electrodes whose moment $\mathbf{M}$ point to different directions.

To understand the magnetic tunnel junction, we first discuss the conductance (G) of finite-sized nanotubes with parallel magnetic moments at the electrodes (*i.e.,* $\theta = 0$), as shown in Fig.1 for (5,5) nanotubes of three different lengths. The behavior of these tubes is strikingly different from that of the more familiar step-like conductance of perfect nanotubes[6,8]. Here, the SWCN tunnel junction shows a resonance behavior, with G sharply peaked at energies where the nanotube has a transmissive level. Note that both the positions and heights of the resonance peaks are sensitive to the nanotube length. To confirm this resonant behavior, we have plotted both G and the local density of states (LDOS) together. The latter measures the electron dwell time[16,17] inside the nanotube region: when a tunneling electron has an energy which matches that of a scattering state, resonant transmission occurs. As expected, there is excellent correspondance between the peaks of G and the LDOS shown in Fig.1. Physically, this resonance behavior may be attributed to the scattering at the contacts between the SWCN and the FM electrodes, which in our model is included within the self-energy of the Green's function. Such scattering contacts are of course absent in the perfect, infinitely long SWCN previously analyzed[6,8]. We note that evidence for such resonant transmissions have been observed experimentally in SWCN systems[18] and predicted for conventional TMR systems[14,19].

In Fig.2 we present the resistance $R(\theta)$ of a (5,5) SWCN magnetic tunnel junction as a function of the angle $\theta$ between the magnetic moments of the FM electrodes. Different panels correspond to different junction lengths and coupling parameters, all having the energy fixed at the Fermi level ($E = 0$). In all cases, a clear spin-valve effect is observed[20] such that $R(\theta)$ varies smoothly with $\theta$. In agreement with the TMR experimental results[1,11], our $FM/SWCN/FM$ device has a minimum resistance at $\theta = 0$, *i.e.,* when the magnetic moments are parallel; and maximum resistance at $\theta = \pi$, *i.e.,* when they are anti-parallel. This variation of the resistance is due to the difference in the parameters $\Gamma_\uparrow$ and $\Gamma_\downarrow$, which in turn reflect the differences in the majority and minority carrier concentration of the FM material[10].

There are, however, several aspects of this spin-valve effect that are unique to SWCN systems. Fig. (2a,b) show that very similar values of $R(\theta)$ are obtained for nanotubes having lengths of 5− and 6−unit cells[12]. However, Fig. (2c) shows that a nanotube having a length of 7− unit cells has a resistance that is one order of magnitude smaller than that of the other two cases! This may be explained by noting that the two shorter junctions are off-resonance at $E = 0$ so that $R(\theta = 0) \gg h/2e^2$, while the longer junction is on-resonance so that $R(\theta = 0) \simeq h/4e^2$ (see Fig.1). What is the reason that a 7-unit cell device is so different from a 6-unit cell device, while the 6-cell device is similar to the 5-cell device? In fact, we have numerically checked all the way up to $N = 58$ and for $N = 99$ and 100 (i.e, SWCN length $\simeq 246\AA$) that all $3N+1$-unit cell devices have small resistance for $\theta = 0$ at the Fermi level. This result can be understood by investigating energy levels of the scattering states of the finite-sized SWCN of the device. For relatively small coupling strength $\Gamma$, the *scattering states* are expected to be close to the *eigenstates* of the corresponding isolated SWCN[17]. It has recently been shown[21], that for an isolated SWCN of finite length, the gap of eigenvalues near the Fermi level oscillates between large and small values as the length is increased. The SWCN magnetic tunnel junctions with lengths $3N+1$ studied here correspond to those cases where the gap is small[21,22]. Thus, when coupled to the device electrodes, which also adds a finite width to the levels, these nanotubes have two scattering states which cross at the Fermi level, and thereby lead to a large conductance or a small resistance at $\theta = 0$.

Hence, if the important device characteristics call for a large current, then SWCN having lengths of $3N+1$ unit cells should be used. Fig.2 also explores the relative importance of the coupling constant $\Gamma_\sigma$. The results suggest that, for a given ratio $\Gamma_\uparrow/\Gamma_\downarrow$, the resistance is relatively less sensitive to values of $\Gamma_\sigma$ on-resonance (Fig.2c), rather than off-resonance (Fig.2a,b). The off-resonance data shows, that a larger $\Gamma_\sigma$ gives rise to a smaller resistance since the SWCN is now contacted better by the electrodes. On the other hand, the on-resonance data shows just the opposite trends because changing $\Gamma_\sigma$ shifts the resonance point slightly, thereby reducing the resonance transmission and therefore increasing the resistance. Finally we note that the off-resonance resistance of Fig. (2a,b) is precisely in the range of 40 to $280K\Omega$, in agreement with the reported experimental values[1].

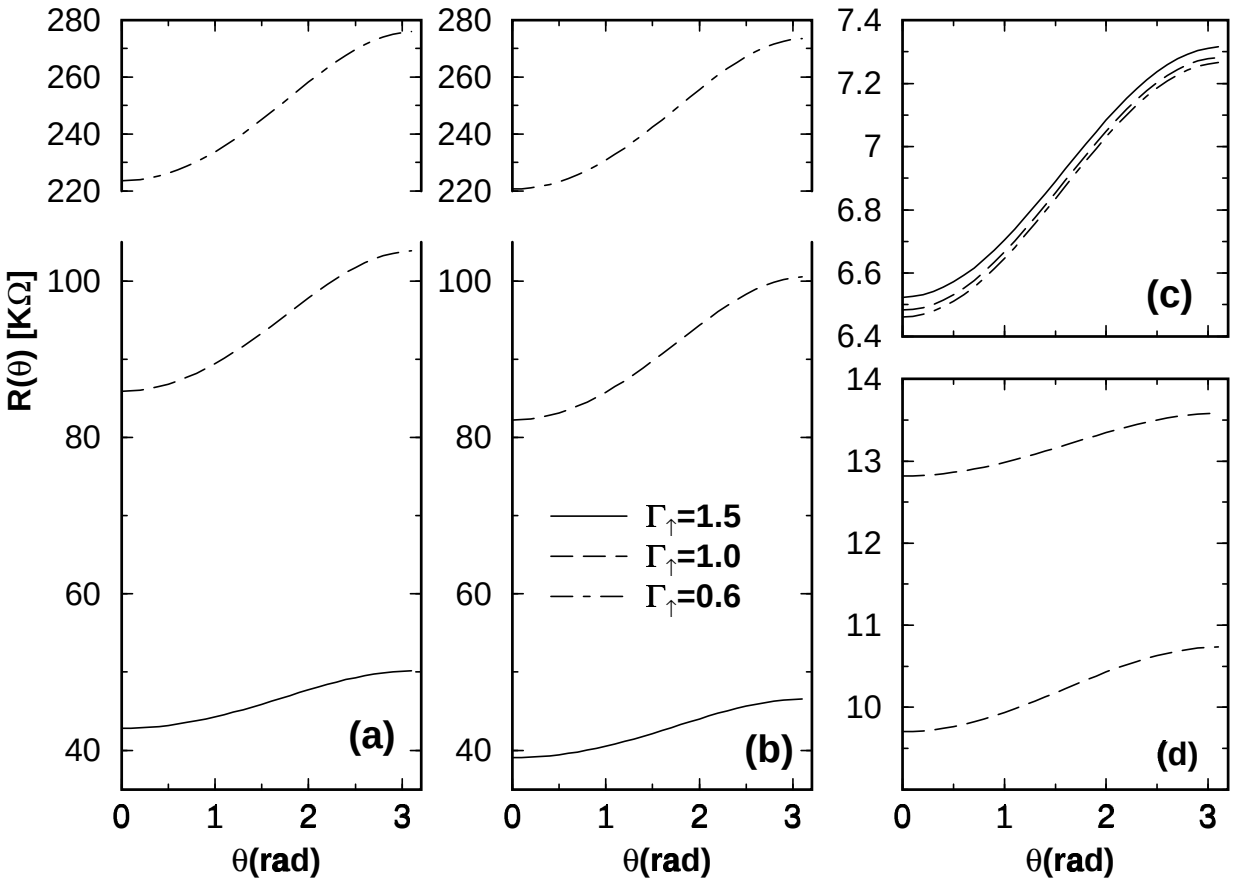

FIG. 2. Conductance of (5,5) SWCN device for different coupling parameters and tube lengths, as a function of the angle $\theta$, showing a spin valve effect. We fixed $\Gamma_\uparrow/\Gamma_\downarrow = 2.0$. (a) for length N=5; (b) for N=6; and (c) for N=7. In (d), we show results for defective (10,10) SWCN -a (7-5-5-7) defect- (lower curve) and -a (5-7-7-5) defect- (upper curve).

Another important quantity of interest is the magneto-conductance ratio $\Delta G \equiv [G(0) - G(\pi)]/G(0)$. The following rough estimates provide insight into the variation of this quantity. From the point of view of tunneling through two barriers, one expects that the contribution to the total current by the up- and down-spin electrons to be: $I_\sigma \propto \Gamma_\sigma^R \Gamma_\sigma^L$ for $\theta = 0$; and $I_\sigma \propto \Gamma_\sigma^R \Gamma_{\bar\sigma}^L$ for $\theta = \pi$. Since the right and the left electrodes are identical except for the orientation of moments, $\Gamma^R_{\uparrow(\downarrow)} = \Gamma^L_{\uparrow(\downarrow)} \equiv \Gamma_{\uparrow(\downarrow)}$, one expects that $G(\theta = 0) \propto \Gamma_\uparrow^2 + \Gamma_\downarrow^2$ and $G(\theta=\pi) \propto 2(\Gamma_\uparrow\Gamma_\downarrow)$, therefore

$$\Delta G = (\Gamma_\uparrow - \Gamma_\downarrow)^2/(\Gamma_\uparrow^2 + \Gamma_\downarrow^2). \tag{5}$$

According to this estimate, $\Delta G$ should be a constant once the ratio $\Gamma_\uparrow/\Gamma_\downarrow$ is fixed. Fig. (3a) illustrates the expected linear dependence of $G(\theta = 0)$ on $\Gamma_\uparrow^2 + \Gamma_\downarrow^2$; Fig. (3b) shows a similar linear dependence of $G(\theta = \pi)$ on $\Gamma_\uparrow\Gamma_\downarrow$. Again, the data for 5- and 6-unit cell devices is increasing due to a stronger coupling to the electrodes; it is decreasing for the 7-unit cell device due to the shifting of the resonance level. Fig. (3c) plots $\Delta G$ versus $\Gamma_\uparrow$ fixing $\Gamma_\uparrow/\Gamma_\downarrow = 2.0$. Our data suggests that $\Delta G$ is indeed roughly constant for the on-resonance case, *i.e.* for the 7-unit cell device. This is consistent with Fig. (2c) where relatively weak dependence on $\Gamma_\sigma$ was found for the resistance. On the other hand, $\Delta G$ is strongly dependent on the parameters for the off-resonance devices. Because of the electronic structure of the SWCN, the magneto-conductance ratio is therefore expected to behave quite differently depending upon whether the length of the nanotube is commensurate with the $3N+1$ rule.

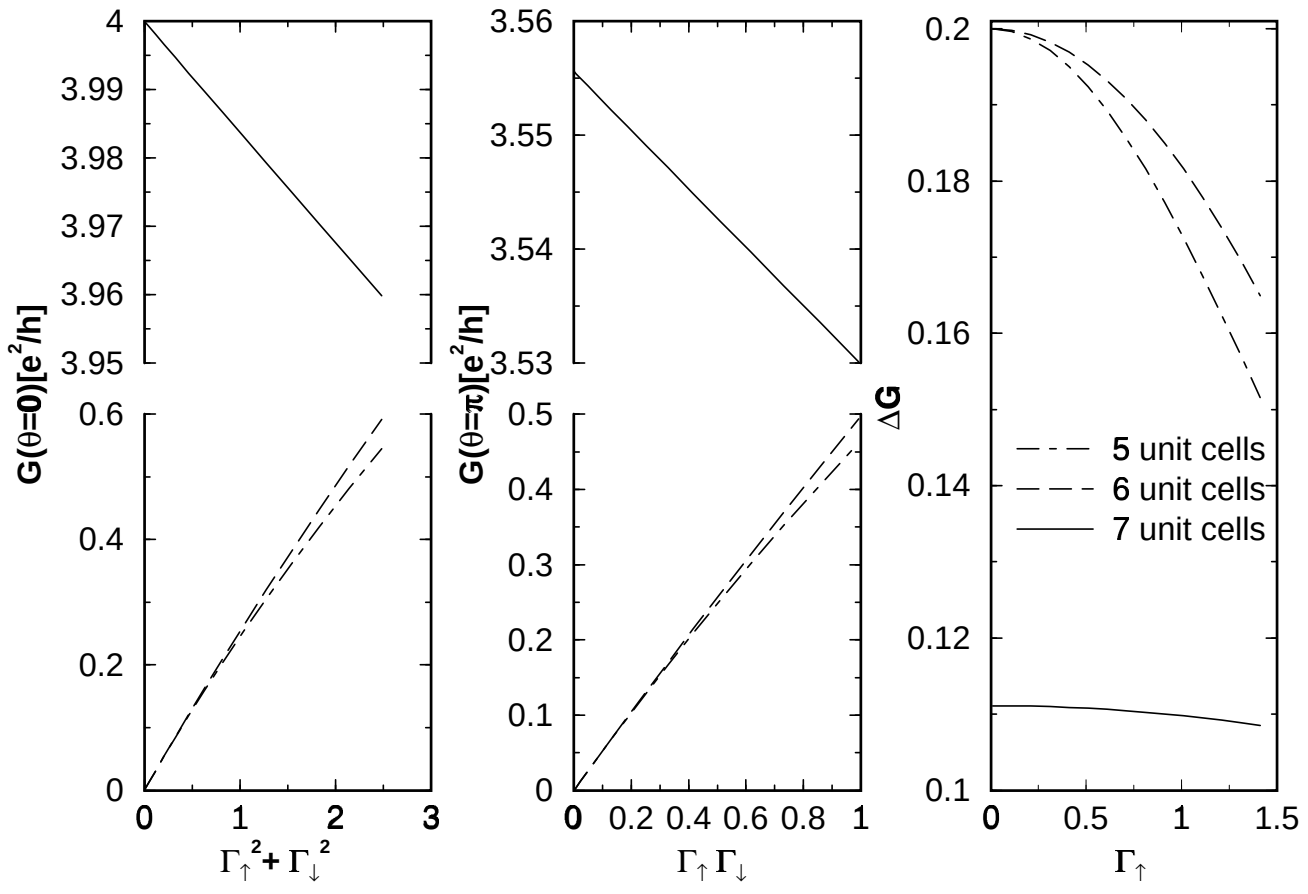

FIG. 3. (a). Conductance at $\theta = 0$ versus $\Gamma_\uparrow^2 + \Gamma_\downarrow^2$ for different nanotube lengths; (b) conductance at $\theta = \pi$ versus $\Gamma_\uparrow\Gamma_\downarrow$; (c) Magneto-conductance ratio $\Delta G$ corresponding to the three devices.

The spin-dependent transport properties discussed above are found to be quite general. Aside from the much longer (5,5) tubes we have also studied (6,6) and (10,10) tubes. Essentially, the same results were obtained. We have also investigated tubes with defects and they have shown similar TMR behavior. Fig. (2d) shows some further examples of this effect for (10,10) tubes with either a (5-7-7-5) or a (7-5-5-7) defect. The former defect forms spontaneously on nanotubes under a large tension via a Stone-Wales transformation, and dominates the initial, mechanical response of the tubes[23]. The latter defect forms in the presence of addimers on strained nanotubes, and represents the first of a set of transformations that can ulitimately lead to formation of a quantum dot[24]. Of course, the presence of these (and other) defects shifts the position of the transmission resonances and provides further scattering to the carriers, so that the $3N+1$ rule is no longer expected to apply. A detailed investigation of the spin-dependent properties of defective nanotubes will be presented in a future publication[25].

In summary, we have shown that there are major dif-ferences in the quantum transport when SWCN are contacted to FM electrodes. First, in the absence of any magnetic effects, transport is characterized by resonant tunneling behavior. Second, a clear spin-valve effect is observed, which enables one to vary the conductance and current continuously through the device as a function of the relative orientation of the magnetic moments of the electrodes. Experimentally, it is possible to control this orientation with an external magnetic field, so that in principle, one should be able to construct SWCN-based spin-valve transistors as already achieved with conventional multilayer technology[26]. Our results further suggest that the transport properties of these devices is sensitive to the length of the nanotubes. In particular, for armchair nanotubes, if the length is commensurate with $3N+1$-unit cells, a resonance behavior leads to a resistance that at least an order of magnitude smaller than those of other tubes. These predictions may be tested experimentally with a two-probe scanning tunneling microscope (STM) in which the tips are constructed out of FM materials. Another important and exciting direction is the possibility of using the present nanotube TMR setup for probing the physics of spin injection into Luttinger Liquid[27] (LL) provided by a nanotube[28]. While we do not expect the spin-valve behavior and the length dependence to alter qualitatively because these effects were due to spin dependent scattering at the contacts and the nanotube molecular orbitals, the spin current is certainly quite distinct if the electron electron interactions are strong. Hence by carefully measuring the temperature dependence of spin current[27], the nanotube TMR may well provide a very good system for the investigations of LL properties.

**Acknowledgments:** H.G. and J.W. gratefully acknowledge Dr. B.G. Wang for many discussions and contributions on the development of the nonequilibrium Green's function theory. We gratefully acknowledge financial support from NSERC of Canada and FCAR of Quebec (H.G); RGC grant (HKU 7115/98P) from the Hong Kong SAR (J.W.); ONR N00014-98-1-0597 and NASA NAG8-1479 (C.R.).

---